\documentstyle[prb,aps,epsfig,multicol]{revtex}
\begin{document}
\sloppy
\title
{Two-step condensation of lattice bosons}
\author{R. Ramakumar$^{1}$ and A. N. Das$^{2}$}
\address{$^{1}$Department of Physics and Astrophysics, University of Delhi,
Delhi-110007, India} 
\address{$^{2}$Saha Institute of Nuclear Physics,
1/AF Bidhannagar, Kolkata-700064, India} 
\date{29 April 2010}
\maketitle
\begin{abstract}
We present a theoretical study of Bose-Einstein condensation in
highly anisotropic harmonic traps. The bosons are considered to
be moving in an optical lattice in an overall anisotropic harmonic
confining potential. We find that two-step condensation occurs for lattice
bosons at much
reduced harmonic potential anisotropy when compared to the case
of an ideal bose gas in an anisotropic harmonic confinement. We also
show that when the bosons are in an isotropic harmonic confinement
but with highly anisotropic hopping in the optical 
lattice, two-step condensation does not occur. We interpret 
some of our results
using single boson density of energy states corresponding to
the potentials faced by the bosons.
\end{abstract}
\maketitle
\section{Introduction}
\label{sec1}
The experimental observation \cite{cornell,davis,bradley} of 
Bose-Einstein (BE) condensation in confined
bose atom clouds have launched extensive experimental and theoretical
studies of this phenomenon and the various properties of the condensates
of free bosons\cite{flbose} and lattice 
bosons \cite{dalfovo,legget,pethicksmith,lewenstein,bloch,yukalov}. 
In three dimensional non-interacting 
free \cite{pethicksmith} or lattice bose systems \cite{ramdas} in 
isotropic harmonic traps, the BE condensation 
is accompanied by a peak in the specific heat at the condensation 
temperature. However, not long after the initial discovery of bose condensation
in harmonic traps, Druten and Ketterle (DK) \cite{drutenketterle} found 
that bosons in highly
anisotropic three dimensional (3D) harmonic traps show 
a qualitatively different behavior 
of the specific heat. Their theoretical calculations showed that 
the peak in the specific heat
of this system 
is not at the bose condensation
temperature, but in a higher temperature range in which 
the bosons, as the temperature decreases, are progressively 
transferred from the tightly confined dimensions to loosely
confined dimension.
Thus, in this system,
the bose condensation occurs in two steps: first the aforementioned
transfer and second the condensation of all bosons 
in to the overall ground state.
The calculations of DK are for free bosons in anisotropic 3D harmonic traps. 
The earlier \cite{sonin} and the later works \cite{shiokawa} 
are also for free bosons
with anisotropic box and anisotropic 3D harmonic confinements, respectively.
\par
The two-step condensation is a phenomenon special in at least two aspects.
The first aspect is that
the condensation of the bosons into the overall
ground state occurs
without any signature in a thermodynamic property (the specific heat).
The second aspect is that it involves a 
dimensionality cross-over in a higher temperature range.
Both of these aspects are of fundamental importance, and hence
an investigation of this phenomenon for {\em lattice bosons}
should be of significant interest.
Now, when
an additional optical lattice potential is applied to bosons in
a harmonic trap, it is known that significant changes occur in
the single boson energy density
of states\cite{ramdas,quintanilla,ramdas2} (DOS).
How these changes alter the two-step
condensation phenomenon is an issue of fundamental importance and
contemporary relevance.
Further, considering that there is a
dimensionality cross-over involved, it should of interest to know
under what conditions a {\em lattice bosons} system in an
anisotropic harmonic trap is truly one-dimensional
as far as thermodynamic properties are concerned.
Motivated by these considerations, in this paper we present a theoretical
study of the condensation of lattice bosons in anisotropic
harmonic traps.
Among other results,
we show that a two-step condensation occurs for lattice bosons
at anisotropies much smaller than that required for free bosons.
This finding may lead to an experimental study of this
phenomenon in optical lattices.
Further, since it is by now relatively easy to experimentally change the boson
hopping in different spatial directions in an optical lattice,
we also study a case of lattice bosons with anisotropic hopping
in an isotropic harmonic confining potential. One may argue that
this is equivalent to bosons with isotropic hopping in an anisotropic
trap and hence two-step condensation would occur in this case as well.
Our calculations show that two-step condensation does not
occur in this case. Our focus in this paper is the effects
of the optical lattice potential on the two-step condensation,
not the effect of boson-boson interactions. The strength
of boson interactions is controllable in optical lattices\cite{bloch,schori},
and hence one can experimentally reach the ideal gas limit to whcih our results
are applicable. 
\par
This paper is organized as follows. In the next section, we revisit the
system of free bosons in an anisotropic harmonic trap. We complement
the analysis of DK by new features we find in the single
boson density of energy states (DOS). In Sec. II, we analyze two-step
condensation of lattice bosons in anisotropic harmonic traps. In Sec. III,
we study the bose condensation of lattice bosons with anisotropic 
hopping in an isotropic harmonic trap. The conclusions are given in
Sec. IV. 
\section{Two-step condensation of free and lattice bosons}
\label{sect2}
\subsection{Free bosons in an anisotropic harmonic potential}
\label{subsec}
In this section we consider boson atoms in an anisotropic harmonic
confining potential ($K_{x}x^{2}+K_{y}y^{2}+K_{z}z^{2}$). The
single boson energy levels are 
$E(n_x,n_y,n_z)$ = $(n_x+1/2)\hbar\omega_x+
(n_y+1/2)\hbar\omega_y+ (n_z+1/2)\hbar\omega_z$,
where $n_x,n_y,n_z$ = 0, 1, 2...$\infty$, $2\pi\hbar$ is the Plank's constant,
and $\omega_{x,y,z}=\sqrt{2K_{x,y,z}/m}$ in which $m$ is the atomic mass.
We use $K_x \ll K_y=K_z$ so that the 
potentials, from the center of the trap, along the $y$ and $z$ 
directions are much steeper compared
to that along the $x$ direction.   
In order to exhibit the two-step condensation in this system, we
calculate the ground state occupancy ($N_0$), the occupancy in the 
loosely confined direction ($N_{1D}$), and the specific heat by  
determining first the chemical potential ($\mu$) 
from the bosons number equation
$
N=\sum_{n_x=0}^{\infty}\sum_{n_y=0}^{\infty}\sum_{n_z=0}^{\infty}
N(E(n_x,n_y,n_z)),
$
where $N(E)=1/(exp[\beta(E-\mu)]-1)$, $\beta=1/k_{B}T$, $T$ the temperature,   
and then calculating $N_0= N(E(0,0,0))$ and 
$N_{1D}=\sum_{n_x=0}^{\infty}N(E(n_x,0,0))$).
The specific heat is obtained from the temperature 
derivative \cite{ramdas} of the total energy
($
E_{tot}\,=\,\sum_{n_y=0}^{\infty}\sum_{n_z=0}^{\infty}
\sum_{n_x=0}^{\infty}N(E(n_x,n_y,n_z))\,E(n_x,n_y,n_z)$). 
We evaluated $N_0$, $N_{1D}$, and $C_v$ numerically
to obtain exact results.
The results of our calculations for 10000 bosons in harmonic traps of
varying anisotropies are shown in Fig. 1. 
The condensation temperature \cite{ramdas,ketterledruten} 
($T_0$), mentioned in this and other figures, is obtained
by setting  $N_0 =0$ and $\mu =E(0,0,0)$ in the bosons number equation.
For low anisotropy, the peak in $C_v$ is
clearly associated with the growth of the bose condensate fraction ($N_0/N$).
With increasing anisotropy, the peak in $C_v$ becomes 
associated with the dimensionality
cross-over in which bosons are transferred from the 
tightly confined directions ($y$ and $z$) into the loosely 
confined direction($x$). To complement the analysis
of DK, we calculated the single boson energy DOS for these systems. 
First we recall that the analytical approximation to the DOS 
of a boson in a 
D-dimensional anisotropic harmonic trap \cite{pethicksmith}
is
\begin{equation}
D(E)= \frac{E^{D-1}}{(D-1)! \prod_{i=1}^{D} (\hbar \omega_{i})}.
\end{equation}
From exact numerical calculations, we find that
the DOS for highly anisotropic traps, shown in Fig. 2a, show features 
qualitatively different from the analytical approximation.
The DOS is found to have one dimensional (1D) 
character (flat regions). In the very
low temperature range, only the lowest flat region is occupied by bosons so
that the $C_v$ shows 1D character as shown in the inset of Fig. 1.
This is consistent with the higher temperature cross-over in which
the bosons are transferred from the tightly confined directions
to the loosely confined direction. We also note that 1D features are
absent in the DOS when the anisotropy is comparatively smaller 
as shown in Fig. 2b. 
In the next section, we consider
consider two-step condensation of lattice bosons.
\subsection{Lattice bosons in an anisotropic harmonic potential}
\label{subsec}
In this section we consider boson atoms in a three dimensional simple cubic ({\em sc})
lattice in an overall anisotropic harmonic 
confining potential. The system Hamiltonian is 
\begin{equation}
H=-t \sum_{<ij>}\left(c^{\dag}_{i}c_{j}+c^{\dag}_{j}c_{i}\right)
  +\sum_{i}V(i)n_{i}-\mu\sum_{i}n_{i},
\end{equation}
where $t$ is the kinetic energy gain when a boson hop
from site $i$ to its nearest neighbor site $j$ in the optical lattice,
$c^{\dag}_{i}$ is a boson creation operator,
$V(i)=(K_{x}x_{i}^2+K_{y}y_{i}^2+K_{z}z_{i}^2)$
is the potential at site $i$, $n_{i}=c^{\dag}_{i}c_{i}$
the boson number operator, and $\mu$ the chemical potential.
We numerically diagonalize this hamiltonian to obtain the
energy levels ($E_l$) of a lattice boson for different values of
the anisotropy of the harmonic trap with $k_x \ll k_y=k_z$, 
where $k_{\alpha} = K_{\alpha} a^{2}$, $\alpha=x,y,z$ and 
$a$ is the lattice constant of the {\em sc} optical lattice. 
The chemical potential and the boson populations in different energy 
levels are calculated  at a temperature $T$, using
the boson number equation: $N\,=\,\sum_{l=0}^{m}N(E_{l})$,
where $E_0$ and  $E_m$ are the lowest and the highest
single boson energy levels and
$N(E_{l})=1/[exp{[(E_{l}-\mu)/k_{B}T]}-1]$.
The specific heat is calculated
from the temperature derivative of
total energy ($E_{tot}=\sum_{l=0}^{m}N(E_{l})E_{l}$).
The lattice size, we use, is large enough that 
the results are free of finite size effects.
The results for the temperature dependence of the
bose condensate fraction ($N_{0}/N$), fractional
boson population in the $x$-direction ($N_{1D}/N$), and
the specific heat for lattice bosons in harmonic traps
with different anisotropies are shown in Fig. 3.
The two-step condensation is clearly seen in highly
anisotropic traps (dash-dot and dotted lines), while
it is absent for small anisotropies. Further, on 
comparing Figs. 1 and 3, we can immediately infer that the
the two-step condensation of lattice bosons is possible for much 
smaller anisotropies compared to the free bosons in
anisotropic harmonic traps. We also 
find that most of the lattice bosons
are confined in the $x$-direction at much higher
temperatures compared to the case of free bosons
even though the anisotropy is much smaller for the former
compared to the latter. This is clear if we compare
$N_{1D}/N$ in both cases, as shown in Figs. 1 and 3, 
for $T > 2.5T_0$. In order to get 
some insight into the two-step condensation occurring
for smaller anisotropies for lattice bosons, we 
calculated the single boson energy DOS for these systems.
The results of our calculations are shown in Fig. 4.
With increasing anisotropy there is a dramatic change
in the single particle DOS (Fig. 4).
Except for very low energies, the DOS for low anisotropy
is very similar to that for the isotropic case.
With increasing anisotropy, the DOS develops features
similar to that expected in 1D. It has peaks separated
by broad regions of comparatively very small values
of DOS. Since the DOS shows strong 1D features for
small anisotropies compared to the free boson case
(without the lattice potential), the 1D behavior in
$N_{1D}/N$ sets in for comparatively smaller
anisotropies of the harmonic traps.
\subsection{Dimensionality cross-overs of lattice bosons}
\label{subsec}
In this section, we investigate the dimensionality cross-over
in some more detail. The specific heat, being a very sensitive 
function of dimensionality changes, is an useful property to study for
this purpose.
As mentioned earlier, for bosons in a highly anisotropic 
harmonic trap, the peak in the specific heat occurs
in a temperature range in which the bosons are transferred
from the tightly confined directions to the loosely confined
direction (here $x$-direction passing through the center of the trap).
With lowering of temperature, when the transfer to the single 
1D chain along the $x$-direction is complete, the system would show
1D behavior. To visualize this cross-over, we have plotted the 
specific heats of lattice bosons in 3D anisotropic traps and 
in a 1D trap in Fig. 5. It is seen that at low temperatures 
the specific heats of the anisotropic systems completely coincide with
that of the 1D chain showing that the anisotropic 3D systems have
become purely 1D. The departure from the 1D behavior starts at
a temperature which is higher for larger anisotropy (Fig. 5).
However, this temperature is much lower than the temperature where 
the specific heat shows the peak and is also lower than the
condensation temperature ($T_0$) for anisotropy as high as 
$k_y/k_x=2\times10^{5}$ for lattice bosons (see Fig. 5). 
It is also found that above this temperature (i.e., where the 
departure from the 1D behavior starts in $C_v$), 
a very small change in $N_{1D}$ produces a very large change in
$C_v$. This is also clear from Fig.1 (for free bosons) and
from Figs. 3 and 5 (for lattice bosons) 
on comparing the temperature dependences 
of $C_v$ and  $N_{1D}$ in the above mentioned region. 
\section{Lattice bosons with anisotropic hopping 
in an isotropic harmonic potential}
\label{sec}
In this section, we investigate if the properties
of lattice bosons with anisotropic hopping
in an isotropic harmonic trap is similar to 
those with isotropic hopping in an anisotropic harmonic trap.
We have studied the effects of anisotropic hopping
on $C_v$ and condensate fraction in isotropic harmonic traps.
The spatial direction dependent hopping we use are such that
$t_x \gg t_y=t_z$.
The results of our calculations for various anisotropy ratios
are shown in Fig. 6. 
We do not find a two-step behavior in this system. The 
specific heat is suppressed with increasing hopping
anisotropy. 
The system shows anisotropic 3D behavior in all the properties studied
even for large hopping anisotropies.
The reason is that even though the hopping is anisotropic, since the
trap is isotropic, one will have a large collection of 
1D chains, symmetric about the central one, coupled by the 
small inter-chain hopping. At the bose condensation temperature,
the bosons condense
in to the central region of the trap.
There is no higher temperature cross-over region as in the
case of lattice bosons with isotropic hopping in a highly
anisotropic harmonic trap.
\section{Conclusions}
\label{sec3}
In this paper we presented a theoretical study of
two-step condensation of lattice bosons in anisotropic
harmonic confining potentials. We find that, even though
significant changes occur in the single bosons DOS, two-step
condensation occurs for lattice bosons and it occurs
for much smaller
anisotropies of the harmonic potential compared to that
for free bosons. This may lead to experimental observation 
of these type of condensations. We find unusual 1D features in the one 
boson DOS both for free and lattice bosons in highly
anisotropic harmonic traps. These DOS's are equally well applicable to
fermions. Using $C_v$ as the indicator,
we also investigated the dimensionality cross-overs in the system.
We find that the system shows purely 1D behavior 
in a much lower temperature range compared
to the temperature range in which the peak
in $C_v$ occurs.
Further, we investigated if two-step condensation occurs
for lattice bosons with anisotropic hopping in an
isotropic harmonic trap. We find that a two-step
condensation does not occur in this system. 
\acknowledgments
RRK thanks Saha Institute of Nuclear Physics for 
hospitality
during part of the time this work was carried out.

\begin{figure}
\includegraphics[width=8in,height=5.5in]{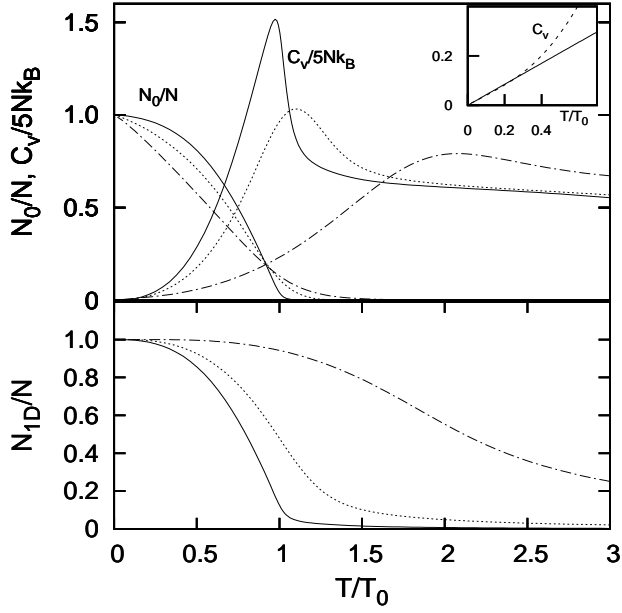}
\caption{
The variation of the specific heat ($C_v$), bose condensate fraction ($N_0/N$),
and the fractional population in the $x$-direction ($N_{1D}/N$) 
with temperature for free bosons in anisotropic harmonic traps. 
$T_0$ is the condensation temperature as defined in the text. 
Solid line: $K_x$ = 0.01, $K_y$ = $K_z$ = 10, 
dotted line: $K_x$ = 0.01, $K_y$ = $K_z$ = 1000, and
dash-dot line: $K_x$ = 0.01, $K_y$ = $K_z$ = 50000.
In our energy units $\hbar\omega_x$ = 0.1414, whereas $\hbar\omega_y$ (= 
$\hbar\omega_z$) = 4.47, 44.72, and 316.2 for
$K_y$ = 10, 1000, and 50000, respectively. 
The number of bosons ($N$) used is 10000. The inset shows
a comparison between $C_v$'s of an anisotropic case (dotted lines
for $K_x$ = 0.01, $K_y$ = $K_z$ = 50000) and a pure 1D case
(solid line) with $K_x$ = 0.01.
}
\label{fig1}
\end{figure}
\begin{figure}
\begin{center}
\includegraphics[width=8in,height=5.5in]{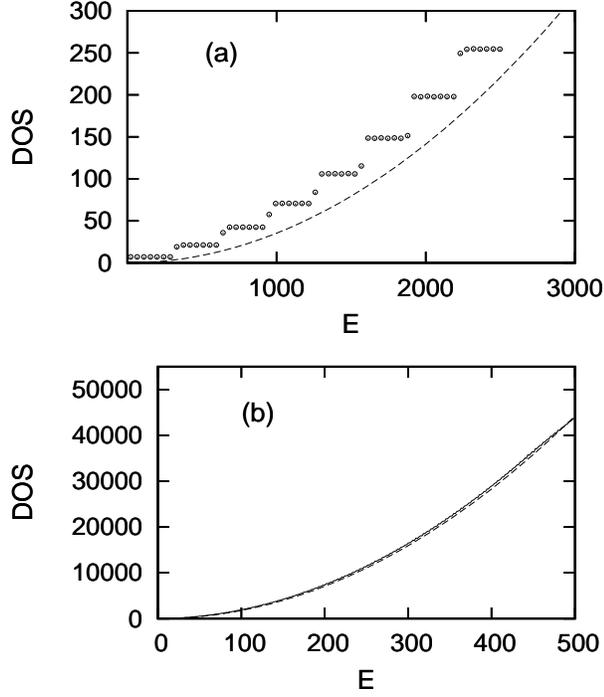}
\caption[]
{
The single particle DOS for free bosons in an anisotropic
harmonic trap vs energy E. Top panel (a): $K_x$ = 0.01 and
$K_y$ = $K_z$ = 50000; triangles give the numerically calculated exact DOS, 
dashed line is the DOS obtained from the analytical expression in Eq. 1.
Bottom panel (b): $K_x$ = 0.01 and  $K_y$ = $K_z$ = 10; thin solid line 
represents numerically calculated exact DOS,  
dashed line is the DOS obtained from Eq. 1.  
In our energy unit $\hbar\omega_x$ = 0.1414.
}
\end{center}
\label{fig2}
\end{figure}
\begin{figure}
\includegraphics[width=8in,height=5.5in]{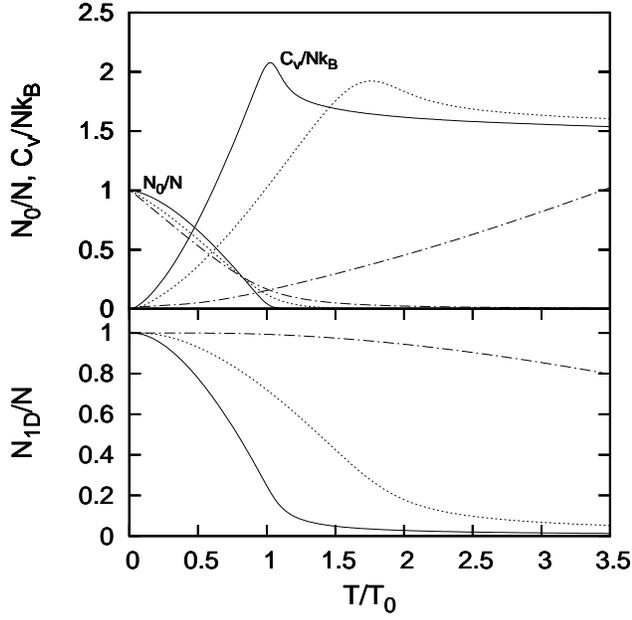}
\caption[]
{
The variation of the specific heat ($C_v$), bose condensate fraction ($N_0/N$),
and the fractional population in the $x$-direction ($N_{1D}/N$) 
with temperature for bosons in combined optical lattice and anisotropic 
harmonic potentials. 
Solid line: $k_x$ = 0.01, $k_y$ = $k_z$ = 10, 
dotted line: $k_x$ = 0.01, $k_y$ = $k_z$ = 100, and
dash-dot line: $k_x$ = 0.01, $k_y$ = $k_z$ = 1000 (in units of $t$ = $1$).
The number of bosons ($N$) used is 10000.
}
\label{fig3}
\end{figure}
\begin{figure}
\includegraphics[width=6in,height=5.5in]{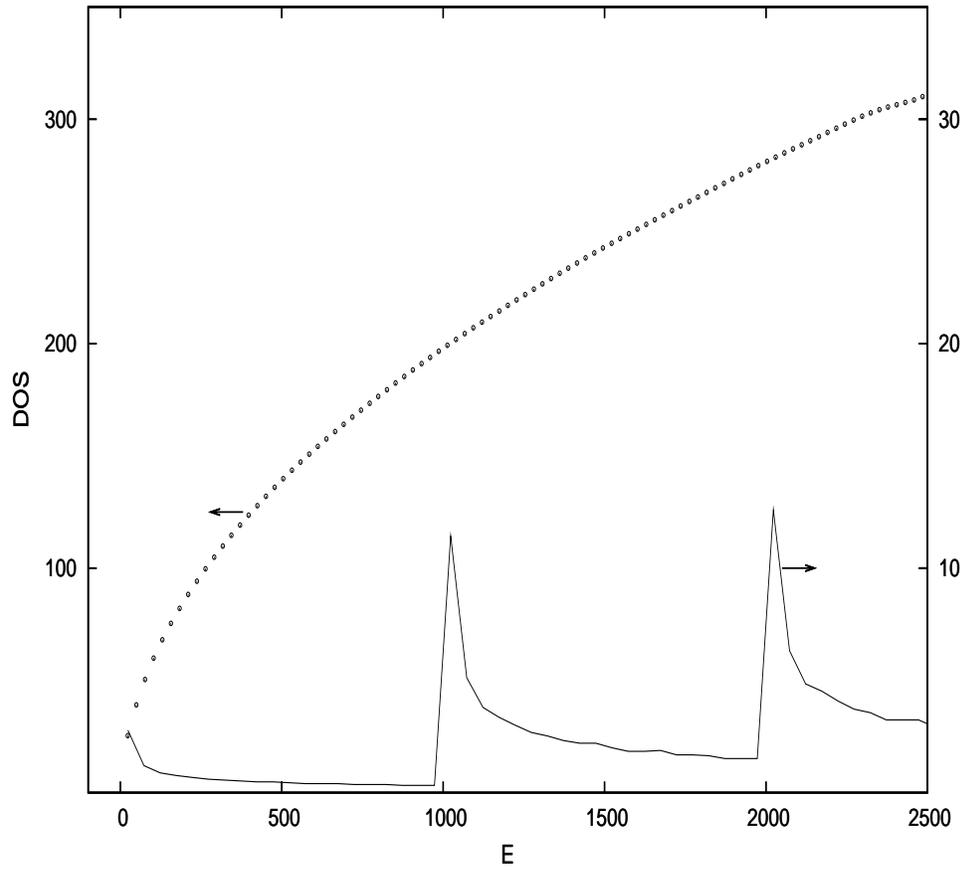}
\caption[]
{
The single particle DOS for bosons in combined optical lattice ($t=1$)
and different anisotropic harmonic potentials.
Circles: $k_x$ = 0.01, $k_y$ = $k_z$ = 10 and
solid line: $k_x$ = 0.01, $k_y$ = $k_z$ = 1000.
}
\label{fig4}
\end{figure}
\begin{figure}
\includegraphics[width=8in,height=5.5in]{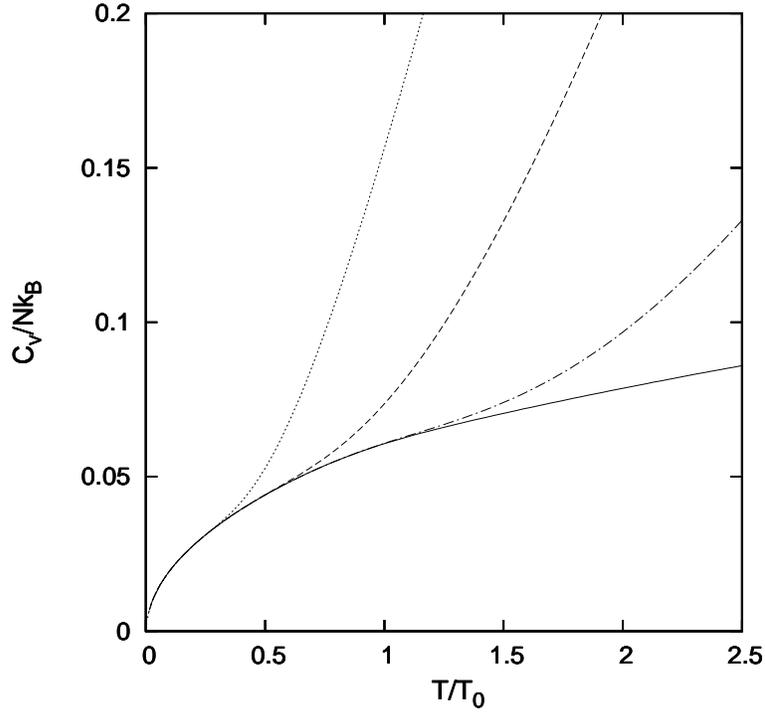}
\caption[]
{ Comparison of the specific heat ($C_v$) of 10000 bosons in the 
3D anisotropic harmonic and optical lattice potentials 
with that of bosons in a 1D combined harmonic and optical lattice. 
Dotted line: $k_x$ = 0.01, $k_y$ = $k_z$ = 1000, 
dash line: $k_x$ = 0.01, $k_y$ = $k_z$ = 2000, and
dash-dot line: $k_x$ = 0.01, $k_y$ = $k_z$ = 4000 (in units of $t$ = 1).
The solid line is for a 1D lattice with $k_x$ = 0.01. 
}
\label{fig5}
\end{figure}
\begin{figure}
\includegraphics[width=8in,height=5.5in]{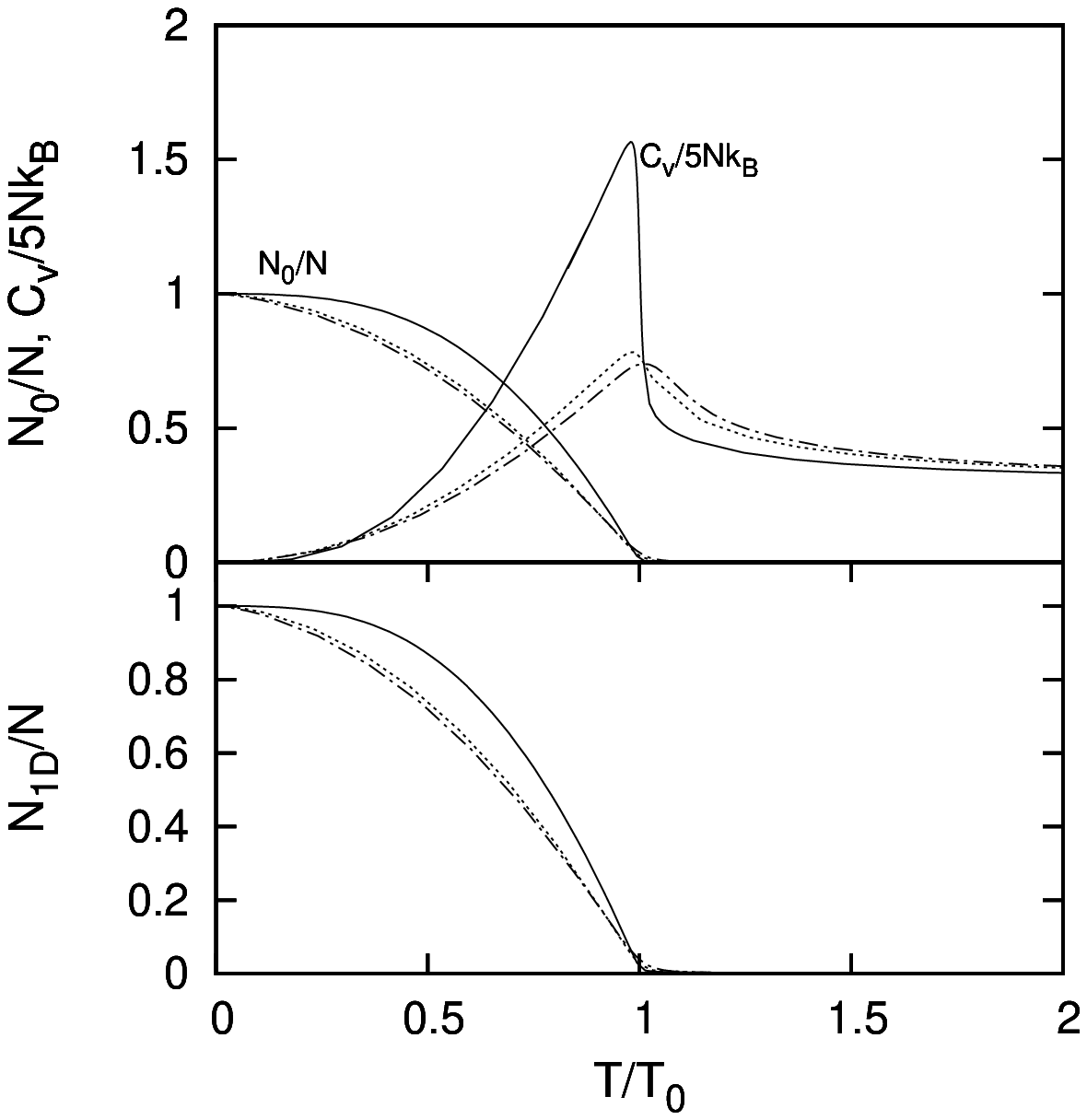}
\caption[]
{
The variation of the specific heat ($C_v$), bose condensate fraction ($N_0/N$),
and the fractional population in the $x$-direction ($N_{1D}/N$)
with temperature for bosons in combined isotropic harmonic ($k_x$ = 
$k_y$ = $k_z$ = 0.01) and anisotropic optical lattice potentials.
Solid line: $t_y$ = $t_z$ = 1.0,
dotted line: $t_y$ = $t_z$ = 0.01, and
dash-dot line: $t_y$ = $t_z$ = 0.001 (all energy parameters are in units 
of $t_x$ = 1.0).
The number of bosons ($N$) used is 10000.
}
\label{fig6}
\end{figure}

\begin{references}
\bibitem{cornell}
M. H. Anderson, J. R. Ensher, M. R. Matthews, C. E. Weiman, and
E. A. Cornell, Science {\bf 269}, 198 (1995).
\bibitem{davis}
K. B. Davis, M. -O. Mewes, M. R. Andrews, N. J. van Druten,
D. S. Durfee, D. M. Kurn, and W. Ketterle,
Phys. Rev. Lett. {\bf 75}, 3969 (1995).
\bibitem{bradley}
C. C. Bradley, C. A. Sackett, J. J. Tollett, and R. G. Hulet, Phys. Rev.
Lett. {\bf 75}, 1687 (1995).
\bibitem{flbose}
The phrases {\em free bosons} and {\em lattice bosons} refers
to bosons in the absence and the presence of an optical 
lattice potential, respectively.
\bibitem{dalfovo}
F. Dalfovo, S. Giorgini, L. P. Pitaevskii, and S. Stringari,
Rev. Mod. Phys. {\bf 71}, 463 (1999).
\bibitem{legget}
A. J. Leggett, Rev. Mod. Phys. {\bf 73}, 307 (2001).
\bibitem{pethicksmith}
C. J. Pethick and H. Smith, {\em Bose-Einstein
condensation in dilute gases} (Cambridge University Press, Cambridge UK, 2002).
\bibitem{lewenstein}
M. Lewenstein, A. Sanpera, V. Ahufinger, B. Damski, A. Sen De, and U. Sen, 
Adv. Phys. {\bf 56}, 243 (2007).
\bibitem{bloch}
I. Bloch, J. Dalibard, and W. Zwerger, Rev. Mod. Phys. {\bf 80},
885 (2008).
\bibitem{yukalov}
V. I. Yukalov, Laser Physics {\bf 19}, 1 (2009). 
\bibitem{ramdas}
R. Ramakumar, A. N. Das, and S. Sil, Eur. Phys. J. D {\bf 42}, 309 (2007).
\bibitem{drutenketterle}
N. J. van Druten and W. Ketterle, Phys. Rev. Lett. {\bf 79}, 549 (1997).
\bibitem{sonin}
E. A. Sonin,   Zh. Eksp. Teor. Fiz. {\bf 56}, 963 (1969) 
[Sov. Phys.-JETP {\bf 29}, 520 (1969)].
\bibitem{shiokawa}
K. Shiokawa, J. Phys. A: Math. Gen. {\bf 33}, 487 (2000).
\bibitem{quintanilla}
C. Hooley and J. Quintanilla, Phys. Rev. Lett. {\bf 93}, 080404 (2004).
\bibitem{ramdas2}
R. Ramakumar and A. N. Das,
Eur. Phys. J. D {\bf 47}, 203 (2008).
\bibitem{schori}
C. Schori, T. St\"{o}ferle, M. Henning, M. K\"{o}hl, T. Esslinger,
Phys. Rev. Lett. {\bf 93}, 240402 (2004).
\bibitem{ketterledruten}
W. Ketterle and N. J. van Druten, Phys. Rev. A {\bf 54}, 656(1996).
\end{references}
\end{document}